\documentclass[twocolumn,showpacs,preprintnumbers,amsmath,amssymb,prd]{revtex4}

\usepackage{graphicx}
\usepackage{dcolumn}
\usepackage{bm}
\usepackage{rotating} 

\newcommand{\E}[1]{ \cdot 10^{#1}}

\begin{document}

\preprint{APS/123-QED}

\title{IGEC2:  A 17-month search for gravitational wave bursts in  2005-2007}

\author{
P. Astone$^1$, L. Baggio$^{2}$, M. Bassan$^{\ast 3,4}$, M. Bignotto$^{5,6}$, 
M. Bonaldi$^{7,8}$, P. Bonifazi$^{1,9}$, G. Cavallari$^{10}$, M. Cerdonio$^{5.6}$, 
E. Coccia$^{3,4}$, L. Conti$^6$, S. D'Antonio$^4$, 
M. di Paolo Emilio$^{11,12}$, M. Drago$^{5,6}$, V. Fafone$^{3,4}$, 
P. Falferi$^{7,8}$, S. Foffa$^{13}$, P. Fortini$^{14}$, 
S. Frasca$^{1,15}$, G. Giordano$^{16}$, W.O. Hamilton$^{17}$, 
J. Hanson$^{17}$, W.W. Johnson$^{17}$, N. Liguori$^{5,6}$, 
S. Longo$^6$, M. Maggiore$^{13}$, F. Marin$^{19,20}$, A. Marini$^{16}$, 
M. P. McHugh$^{21}$, R. Mezzena$^{8,22}$, P. Miller$^{17}$, 
Y. Minenkov$^{11}$, A. Mion$^{8,22}$, G. Modestino$^{16}$, A. Moleti$^{3,4}$, 
D. Nettles$^{17}$, A. Ortolan$^{18}$, G.V. Pallottino$^{1,15}$,
G. Pizzella$^{16}$, S. Poggi$^{23}$, G.A. Prodi$^{8,22}$, V. Re$^{8,22}$, 
A. Rocchi$^4$, F. Ronga$^{16}$, F. Salemi$^{24}$, R. Sturani$^{13}$, 
L. Taffarello$^{5}$, R. Terenzi$^4$, G. Vedovato$^6$, 
A. Vinante$^{25}$,, M. Visco$^{3,9}$, S. Vitale$^{8,22}$
J. Weaver$^{17}$, J.P. Zendri$^{6}$ and P. Zhang$^{17}$
}

\address{$^1$ INFN, Sezione di Roma, P.le A.Moro 2, I-00185, Roma, Italy}
\address{$^2$ CNRS, LATMOS, 78280 Guyancourt, France}
\address{$^3$ Dipartimento di Fisica, Universit\`a di Roma ``Tor Vergata'', Via Ricerca Scientifica 1,  I-00133 Roma, Italy}
\address{$^4$ INFN, Sezione di Roma Tor Vergata, Via Ricerca Scientifica 1, I-00133 Roma, Italy}
\address{$^5$ Dipartimento di Fisica, Universit\`a di Padova, Via Marzolo 8, 35131 Padova, Italy}
\address{$^6$ INFN, Sezione di Padova, Via Marzolo 8, 35131 Padova, Italy}
\address{$^7$ Istituto di Fotonica e Nanotecnologie, CNR-Fondazione Bruno Kessler, 38123 Povo, Trento, Italy}
\address{$^8$ INFN, Gruppo Collegato di Trento, Sezione di Padova, I-38050 Povo, Trento, Italy}
\address{$^9$ INAF, Istituto Fisica Spazio Interplanetario, Via Fosso del Cavaliere, I-00133 Roma, Italy}
\address{$^{10}$ CERN, Geneva - Switzerlad}
\address{$^{11}$ INFN, Laboratori Nazionali del Gran Sasso, Assergi, L'Aquila,Italy}
\address{$^{12}$ Dipartimento di Fisica, Universit\`a de L'Aquila, L'Aquila, Italy}
\address{$^{13}$ D$\acute{e}$partement de Physique Th$\acute{e}$orique, Universit$\acute{e}$ de Gen\`eve, Gen\`eve, Switzerland }
\address{$^{14}$ Dipartimento di Fisica, Universit\`a di Ferrara and INFN, Sezione di Ferrara, I-44100 Ferrara, Italy}
\address{$^{15}$ Dipartimento di Fisica, Universit\`a di Roma ``La Sapienza'', P.le A.Moro 2, I-00185, Roma, Italy}
\address{$^{16}$ INFN, Laboratori Nazionali di Frascati, Via E.Fermi 40, I-00044, Frascati, Italy}
\address{$^{17}$ Department of Physics and Astronomy, Louisiana State University, Baton Rouge, Louisiana 70803}
\address{$^{18}$ INFN, Laboratori Nazionali di Legnaro, 35020 Legnaro, Padova, Italy}
\address{$^{19}$ INFN, Sezione di Firenze, I-50121 Firenze, Italy}
\address{$^{20}$ LENS and Dipartimento di Fisica, Universit\`a di Firenze, I-50121 Firenze, Italy}
\address{$^{21}$ Department of Physics, Loyola University New Orleans, New Orleans, LA 70118,USA}
\address{$^{22}$ Dipartimento di Fisica, Universit\`a di Trento, I-38050 Povo, Trento, Italy}
\address{$^{23}$ Consorzio Criospazio Ricerche, I-38050 Povo, Trento, Italy}
\address{$^{24}$MPG, Albert Einstein Institut, 30167 Hannover, Deutschland}
\address{$^{25}$LION, Kamerlingh Onnes Laboratorium, Leiden University, NL-2300 RA Leiden, The Netherlands}

\email[Corresponding author ]{massimo.bassan@roma2.infn.it}

\begin{abstract}

We present here the results of a 515 day search for short burst of gravitational waves by the IGEC2 observatory. This network included 4 cryogenic resonant-bar detectors: AURIGA, EXPLORER and NAUTILUS in Europe, and ALLEGRO in America.
These results cover the time period from Nov 6 2005 until Apr 15 2007,
partly overlapping the first long term observations by the LIGO interferometeric detectors. 
The observatory operated with high duty cycle, namely 57\% for 4-fold coincident observations, and 94\% for 3-fold observations. The sensitivity was the best ever obtained by a bar network: we could detect impulsive events with a burst strain amplitude $h_{rss} \sim 1\E{-19}$ with an efficiency >50\%.
The network data analysis was based on time coincidence searches over
at least three detectors, used a blind search technique and
was tuned to achieve a false alarm rate of 1/century. 
When the blinding was removed, no gravitational wave candidate was found.
\end{abstract}
  
  \maketitle

\section{\label{sec:intro}Introduction}
In the last decade, while interferometric detectors were approaching, through a series of commissioning runs, their target sensitivity, resonant bar detectors continued to  reliably observe the cosmos for very long periods of time, looking for unmodeled impulsive gravitational waveforms (GW) from galactic sources.

For the purpose of coordinating this search effort, the four cryogenic, resonant mass detectors (plus, in an early stage, the Australian antenna Niobe) joined forces in a collaborative agreement called IGEC (International Gravitational Event Collaboration) that mainly consisted of a protocol for data exchange and analysis. The IGEC collaboration analyzed almost 4 years of data from 5 antennas \cite{IGEC1a}\cite{IGEC1b}.
This protocol was later revised and renamed IGEC2: among other changes, it included coordination on scheduling routine maintenance operations (refills of cryogenic fluids) in order to ensure maximal time coverage with at least 3 of the 4 detectors. A first period of 6 months of data was analyzed and published under the new protocol \cite{IGEC2}: it yielded a negative result (no GW candidates) but it was instrumental in setting up and testing network analysis procedures that were the starting point from which the present analysis has evolved.

The detectors taking part in the IGEC2 network are: ALLEGRO, located at Louisiana State University (Louisiana-USA) and operated by the local ALLEGRO group,  AURIGA, located in the Legnaro National Laboratories (Padova-Italy) of INFN and operated by the AURIGA collaboration, EXPLORER, located at CERN (Geneva-Switzerland) and NAUTILUS, located in the Frascati National Laboratories (Frascati-Italy) of INFN. Both latter detectors are operated by the ROG collaboration. AURIGA, EXPLORER and NAUTILUS all benefit from support from INFN, while ALLEGRO was funded by NSF.

This network of four detectors, all operating with high duty cycle and large overlap time, is effective in rejecting spurious candidate events caused by transient local disturbances or by intrinsic detector noise.  The sensitivity of these antennas has been superseded by the much superior performance of large interferometers like LIGO and VIRGO. Therefore, IGEC2 upper limits on the GW flux on Earth are no longer astrophysically significant so in this paper we target the possible detection of a rare, impulsive event with long term observations.

In the previous runs of this network, either the sensitivity of the antennas was lower \cite{IGEC1a}\cite{IGEC1b}, or the observation lasted for a much shorter stretch of time \cite{IGEC2}.
The LIGO observatory has also carried out searches for bursts, both with its three interferometers \cite{LIGOS1}\cite{LIGOS3}\cite{LIGOS4} and in a coincidence run with GEO \cite{LIGOGEOS4}.

The first published results \cite{bursts5, LIGOS5} of the LIGO S5 run report
a relevant advance in these searches, yielding a significant lowering
in the previously published upper limits. That work refers to the
period November 2005 through November 2006, covered with a $\sim 70 \%$
duty cycle. The same period is covered by the present work, clearly
with a lower sensitivity, but with a much larger duty cycle ($\sim 94 \%$).   For this reason it can be used in conjunction with this early phase of S5, when the most common periods were stretches of time with only one interferometer taking data.

Several astrophysical processes can generate gravitational radiation in the sensitive bandwidth of the IGEC2 detectors (all centered around 900 Hz): among these, stellar core collapse \cite{Scheidegger}, final phase of inspiralling and ring down in merger of compact binary systems, as well as rotational instabilities and quasi-normal mode oscillations of  relativistic stars \cite{valeria}  \cite{kostas}. 
The amplitude of these events, according to current estimates \cite{LRR2003} are such that an event in our galaxy (say within a few kpc) can be detected by the resonant antennas.
 In view of the large uncertainty in the modeling of these waveforms and of the restricted bandwidth of our detectors, we search our data for unmodeled, featureless short bursts:
as in the previous search \cite{IGEC2}, we aimed at signals with typical durations up to a few tens of milliseconds (see the analysis of sect \ref{sec:network}), e.g. damped sinusoids, ring down, or gaussian bursts.

The main advances reported here 
refer to a detailed analysis on the choices of the coincidence window and of the thresholds strategy, carried out through an extensive use of software injections of pulses in the data.
In addition, we analyzed a substantial stretch of time of fourfold operation.
This paper describes the results of 17 months of observation, from Nov. 16$^{th}$ 2005 to Apr. 14$^{th}$ 2007, containing the longest reported period of fourfold coincidence observation, 293.5 days out of 515. Indeed, in the first IGEC search, published in 2003 \cite{IGEC1b}, only 26 days of fourfold observation were collected at a much lower sensitivity, in a four year span and using five antennas. In the 2007 paper\cite{IGEC2} we only analyzed threefold coincidences.

The goal of this observation campaign, just as in that reported in \cite{IGEC2} was to look for coincidence events with a background of accidentals of 1 per century. This low level of false alarm was achieved with a suitable choice of the thresholds applied to the candidate event lists produced by each group, as discussed in detail in sect.\ref{sec:network}. We chose to undertake five searches in parallel: one four-fold coincidence search and four triple coincidence searches,  \textit{a priori} assigning to each a false alarm rate (FAR)  of $0.2 /century$. Once these five searches are combined in a logical OR, the desired level of FAR is achieved. Two-fold coincidences were not analyzed, since in order to achieve a reasonable value of the FAR, we should have had to raise the thresholds so much that the efficiency would have been even lower than for the threefolds.

In the following section we review some features of the network, of its detectors and of the procedure used to produce the exchanged data. 

In sect \ref{sec:network} we describe the study carried out to evaluate the background and set up the corresponding cuts in the data that guarantee the desired FAR. Finally the results of the search are presented and discussed in the last section.

\section{\label{sec:data}The IGEC2 observatory: the four detectors and their data}

The four cryogenic resonant GW detectors that are part of the IGEC2 network are reliable and stable machines.  Their long term noise performance is rather stationary. Therefore, both the general observatory description and the noise characterization of the antennas are substantially unchanged with respect to that described in \cite{IGEC2}. While detailed description of the equipment can be found in \cite{Auriga, Auriga2006} for AURIGA, in \cite{ROG,ROGprl}  for EXPLORER and NAUTILUS and in \cite{Allegro} for the ALLEGRO detector, respectively, here we will just recall the common features of the antennas.

These four detectors are very similar in their experimental set up. They all consist of a 3 m long Aluminum bar, suspended in  vacuum in a cryogenic environment. The quantity we monitor and analyze,  in search of an event generated by GW, is the amplitude of vibration of the first longitudinal elastic mode of the cylinder, resonating around 900 Hz. This signal is intrinsically narrow band, the bandwidth being set by the interplay between thermal (resonant) noise and amplifier (wide-band) noise and reaching at best 10\% of the operating frequency. For this reason, the signal we can extract from the antenna vibration can be related to the Fourier amplitude $H(\omega)$ at the resonant frequency of an hypothetical short GW signal, rather than to the wave amplitude $h(t)$ itself.

\begin{figure}
\includegraphics[width=20pc]{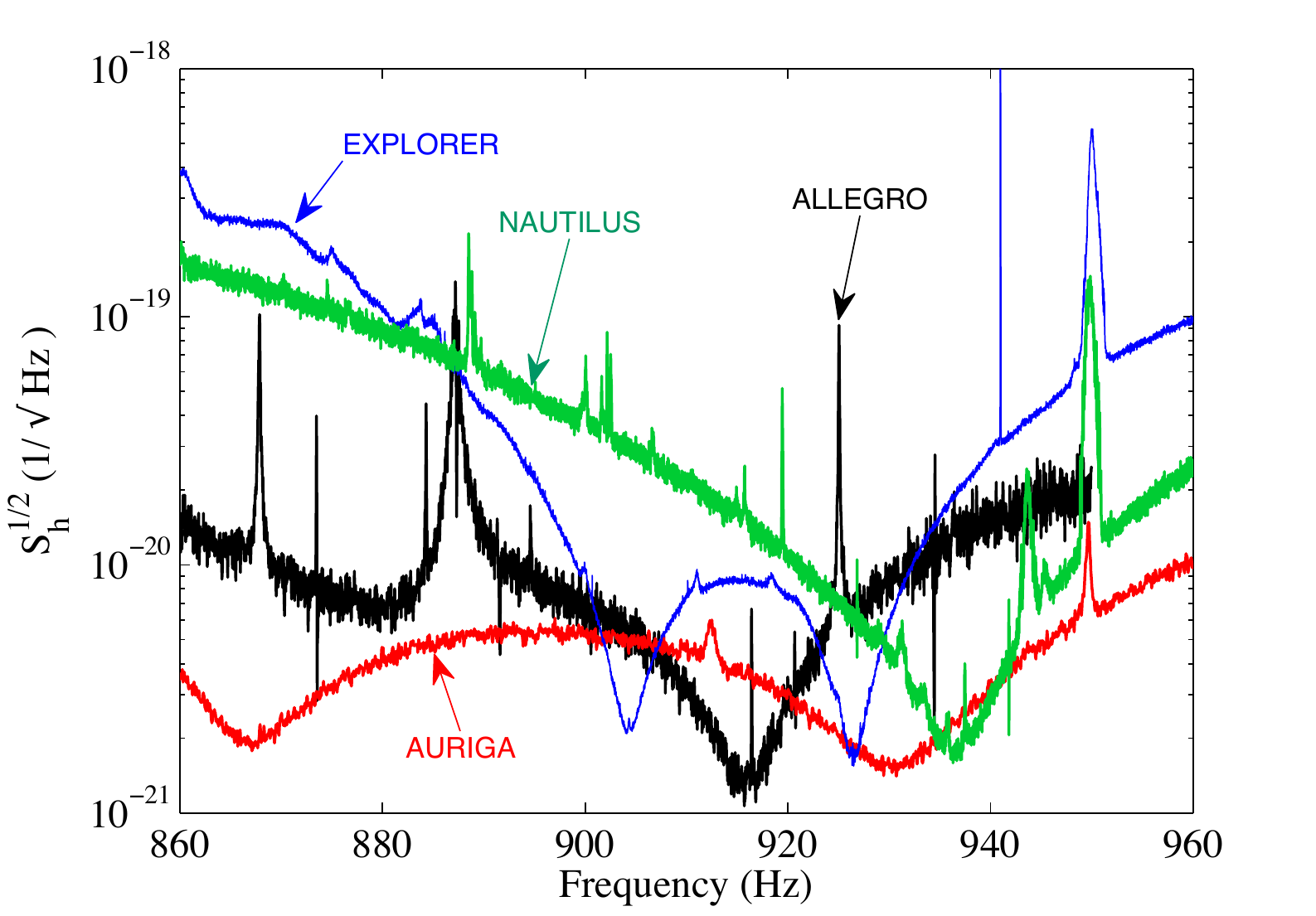}
\caption{Typical strain noise spectral densities (single-sided) of IGEC2 detectors.
All detectors are sensitive in a region around 900 Hz, and the minimum
level of noise is comparable in all spectra. The wider bandwidth of
AURIGA includes the bandwidths of the other detectors. The noise curve of Explorer (blue on-line) 
refers to operation  after the  2006 retuning.}
\label{fig:Shh} 
\end{figure}

The spectral sensitivity curves of the four antennas are shown in fig.\ref{fig:Shh}. Different spectral shapes in the various detectors arise from particular choices of the antenna-readout coupling. Indeed, 
the resonant transducer that is used to convert the bar vibration into an electric signal (a light mass mechanical oscillator, whose vibrations modulate an electric or magnetic field), is different in the four detectors. While ALLEGRO has a superconducting, persistent-current, inductance-modulation device, AURIGA, NAUTILUS and EXPLORER rely on a capacitive, constant charge biased transducer. In all detectors the first stage amplifier is a d.c. SQUID, an ultra-low noise device. AURIGA indeed uses a double d.c. SQUID with the matching LC circuit tuned to the mechanical antenna frequency (resulting in a three mode resonant system). On the other hand, EXPLORER and NAUTILUS, while using a loosely tuned \cite{ROGprl} matching circuit, employ a very small gap capacitor ($8- 12 \mu m$) to achieve a high electromechanical coupling. In ALLEGRO, due to the low source impedance of the inductive transducer, no matching transformer is needed. Both NAUTILUS and AURIGA are equipped with  dilution refrigerators that allow them to operate at ultralow temperature. However, because of reliability and duty cycle considerations, they were in this case operated at about 3 - 4 K, like EXPLORER and ALLEGRO.
With respect to the spectral sensitivity curves shown in \cite{IGEC2},  the $S_{h}(f)$ of Explorer now has a more symmetric shape. This is due to a small change in the bias voltage of the resonant capacitive transducer that was implemented in April 2006.

All four detectors require periodic down-time for cryogenic maintenance, typically one or two days per month for refilling cryogenic fluids.
A coordination effort was made to undertake maintenance duties on different days for each apparatus, in order to maximize the time with at least three detectors simultaneously in operation. As a result, we achieved the observation times described in table \ref{table:giorni}, that shows that we had three or more antennas operating for 94\% of the entire 515 day period with data quality suitable for science goals.

\hspace{1cm}
\begin{table}[h]
	\centering
			\begin{tabular}{|c|c|}\hline
			 \textbf{Configuration} & \textbf{Time of operation }  \\ 
			 & \textbf{(days)}  \\\hline 
			\textbf{0 detector}   &0\\   \hline  
			\textbf{1 detectors}  &1.6\\ \hline  
		  \textbf{2 detectors}  & 31.0 \\ \hline  
			\textbf{3 detectors}  &188.8\\ \hline
			\textbf{4 detectors}&293.5\\ \hline		 	
		\end{tabular}
	\caption{Multiplicity of observing antennas during the analyzed period.}
\label {table:giorni}
\end{table}

\begin{figure}[h]
\includegraphics[width=20pc]{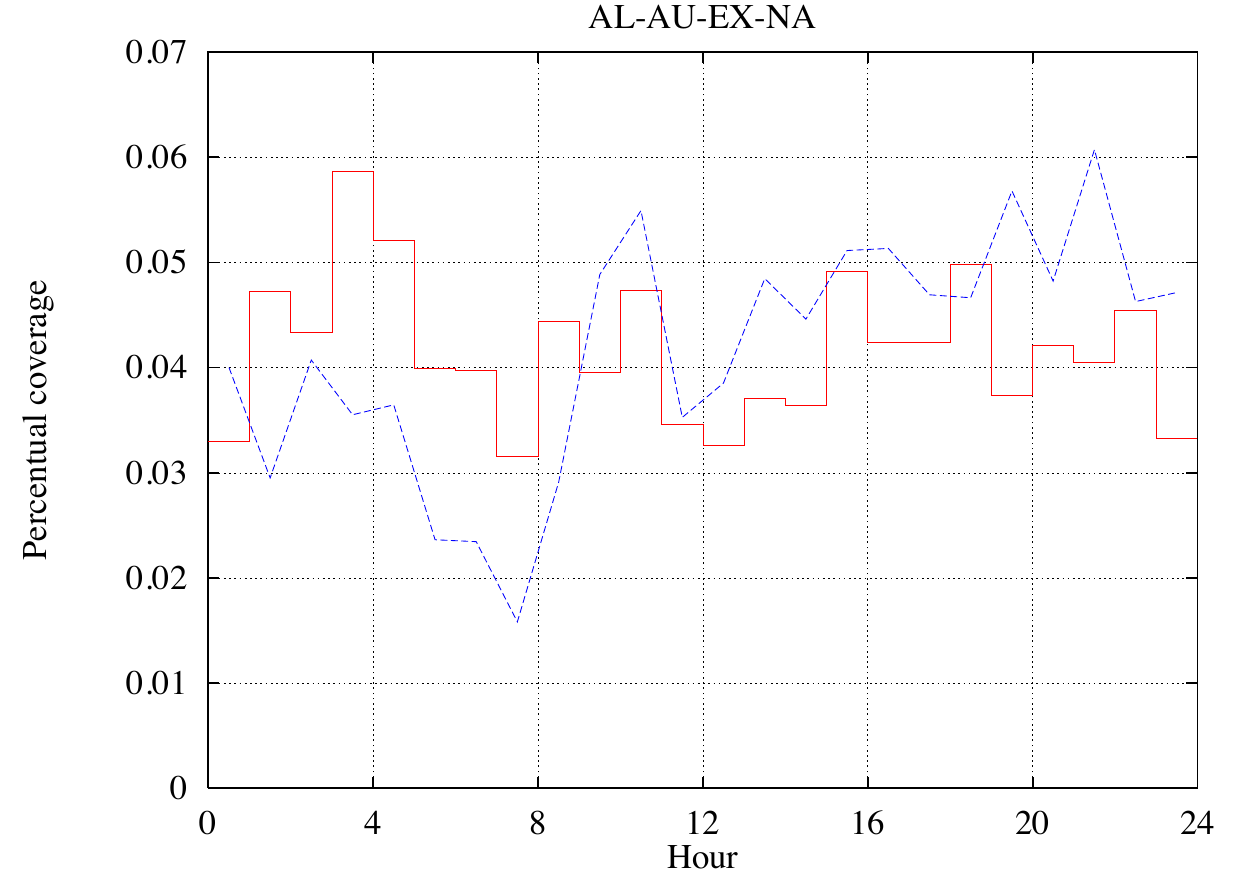}
\caption{Distribution of the fourfold observation time in the hours of the day: sidereal (solid line, red) and solar (dotted line, blue).}
\label{oress}
\end{figure}

This long period of observation and the high duty cycle of all detectors guarantee, despite the periodic maintenance stops, an adequate coverage of all hours in the diurnal time, as shown in fig.\ref{oress}, where the percentage of time of the fourfold mode of operation is plotted vs the hour of the day. While the sidereal time distribution has small fluctuations around the expected value of 4.17\%, the solar time distribution exhibits larger oscillations with a dip in coincidence with the start of local morning activities in Italy and Geneva.

As in previous joint searches \cite{IGEC2}, each group is responsible for the production and the calibration of the data of its own detector. This includes independent design and operation of the antenna hardware, data sampling and filtering and extraction of the candidate events, by proper choice of the amplitude thresholds. All groups filter the data by applying adaptive linear filters matched to $\delta $-like signals. We  carried out calibrations of these filters based both on mechanical excitations with short pulses and on software injections
of various impulsive waveforms. These calibrations show that the
$\delta$-filtered data preserve a good efficiency also to longer signals, such as damped sinusoids with decay times $\tau \lesssim 30$ ms, as it will be discussed in more details in sec.IIIA.

A candidate event is identified by detecting a local maximum in the absolute value of the filtered data stream. The occurrence time of the maximum is the estimate of the arrival time of the burst and its amplitude is the estimate of the Fourier amplitude $H$ of the GW waveform $h(t)$, assumed to be a delta. In order to remove short time clustering in the events of each detector, we introduced a dead time in the process of selection of the events. The dead time represents the minimum time allowed between two different events, and its value was chosen 1 s for EXPLORER, NAUTILUS and AURIGA and 0.2 s for ALLEGRO.

Thus, AURIGA, EXPLORER and NAUTILUS adaptively set their thresholds, depending on the local (in time) noise level. However, due to a different data production process, ALLEGRO, whose data where not used in the 2005 search, produced its own events with a fixed H threshold. 

\begin{figure}[h]
 \includegraphics[width=20pc]{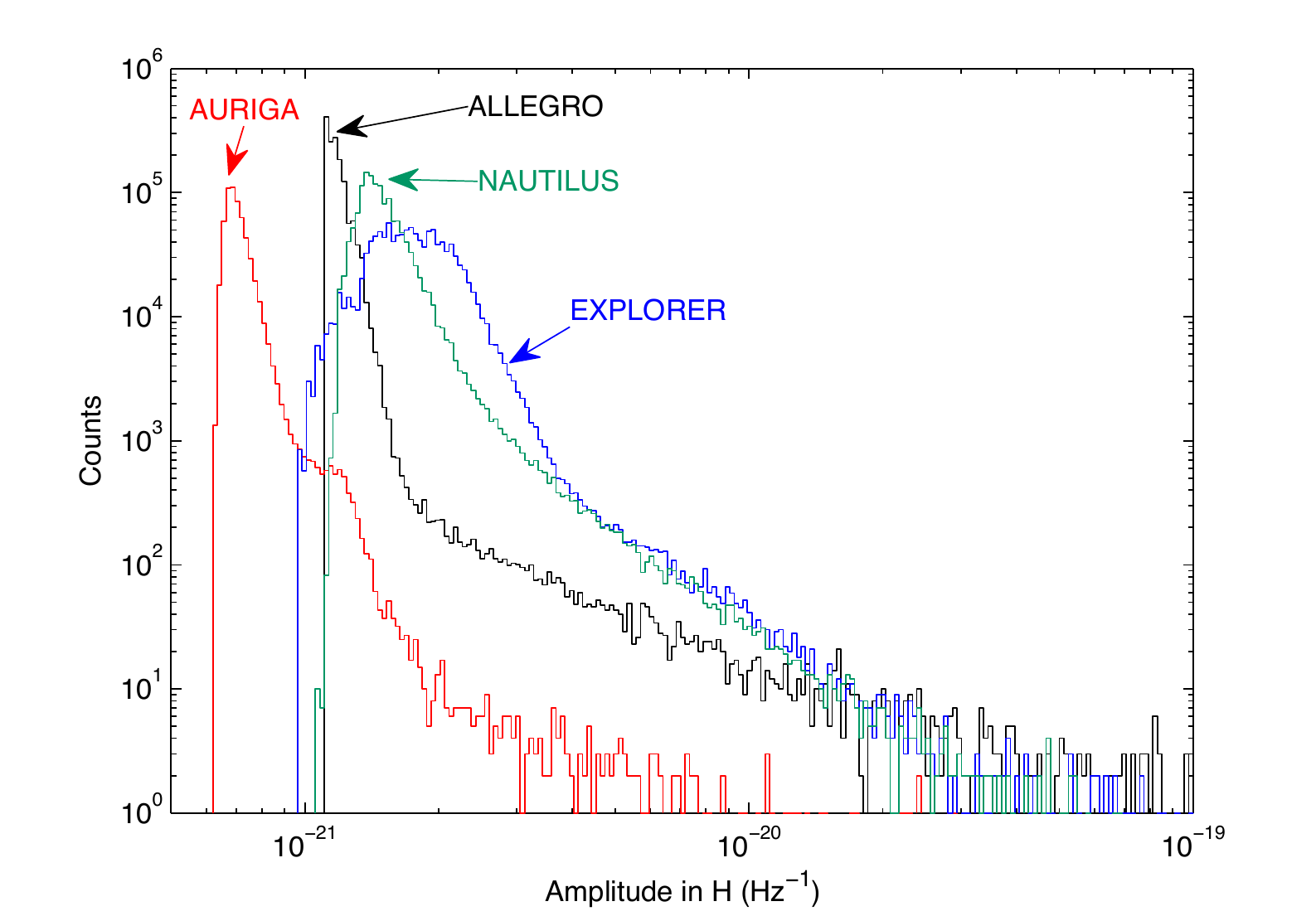}
\caption{Amplitude distribution of the exchanged candidate events above the minimal thresholds (AURIGA $SNR > 4.5$, ALLEGRO $H> 4 \cdot 10^{-21} Hz^{-1}$, EXPLORER and NAUTILUS $SNR > 4.0$).}
\label{fig:histoevents} 
\end{figure}

The SNR (or H for ALLEGRO) threshold value and the resulting number of event candidates, as well as the average noise level of each detector are listed in table \ref{tab:thr}. Note that, although independently chosen by each group, the applied thresholds yield a comparable number of events (1 to 3 thousand per day)  from all detectors. The ampllitude distribution of all exchanged events is shown in fig. \ref{fig:histoevents}.

Each group exchanged the events list after adding a time offset,
kept confidential to the other groups.
Background evaluation and the tuning of the time coincidence analysis were therefore completed in a blind manner; only after agreeing on the thresholds, coincidence windows and other choices that contribute the search strategy were the confidential shifts revealed (\textit{opening the box}), thus comparing the real candidate event times and finding the actual coincidences in the data.

The data exchange and background evaluation was carried out in two installments: first the event candidates for the period Nov 16th 2005 to Dec 31st 2006 (411 days) were exchanged and analyzed. The background was evaluated and  thresholds were set in order to obtain a FAR of 1/century. We call this \textrm{Data Stretch A}. After \textit{opening the box}, 104 more days of data, hence referred to as \textrm{Data Stretch B}, became available. As of April 15th, 2007, ALLEGRO ceased operations. 
\hspace{1cm}
\begin{table}[h]
	\centering
			\begin{tabular}{|c|c|c|c|}\hline
\textbf{Detector}& \textbf{Noise} &\textbf{Threshold} & \textbf{Number} \\ 
 & $(Hz^{-1})$  &  & \textbf{of events} \\\hline 
			\textbf{ALLEGRO}  &$2.7\cdot10^{-22}$ & $H =1.1 \cdot 10^{-21} Hz^{-1}$ &1,472,517\\   \hline  
			\textbf{AURIGA}  &$1.5\cdot10^{-22}$ &SNR=4.5&585,968\\ \hline  
		  \textbf{EXPLORER}  &$4.1\cdot10^{-22}$ &SNR=4& 1,193,830 \\ \hline  
			\textbf{NAUTILUS}  &$3.5\cdot10^{-22}$&SNR=4&1,400,882\\ \hline
					\end{tabular}
	\caption{The average noise level, the value of threshold chosen by the responsible group, and the resulting number of exchanged events, for the four detectors.}
	\label{tab:thr}
\end{table}

We thought it reasonable to consolidate in this paper the entire duration of four-fold operations, but this required a separate analysis for stretch B, as that for stretch A had already been completed, including the exchange of the confidential time shift. In  section \ref{sec:network} we describe the background evaluation carried out separately for the two data stretches.

\section{\label{sec:network}Network Data Analysis}

We ran five separate searches of coincidences: one fourfold,
and four triples. Their observation times are well overlapped and we combined them in a logical OR, to improve the overall detection efficiency.  In order to achieve the desired total FAR of 1/century, we decided to equally partition it into each separate search.

Once each group has produced its list of candidate events, the only parameters that can be adjusted to set the desired number of accidentals, i.e. 0.2/century in each search, are the coincidence window and a set of second (obviously higher) thresholds.

\subsection{The time coincidence window }

In the past \cite{IGEC1b}\cite{IGEC2}, we used approximate analytical models to estimate the time response of any detector to delta-like signals. 

The time coincidence window was then set, with the goal of keeping the efficiency to 95\%, by using the estimated time response within the Tchebichev inequality.

For this analysis however, we were able to take advantage of an extensive methodological study jointly carried out with the VIRGO group \cite{virgobarre}. In this study, we applied, to one 2005 day of real data, a large number of software injections representing damped sine waves 
(a very convenient and widely used representation of short bursts) 
 with various center frequencies (between 866 and 946 Hz) and various durations ($\tau$ = 1, 3, 10 and 30 ms):
\begin{equation}
h(t) = h_0 sin(2 \pi f_0 t) e^{-t/\tau}   \hskip  1cm  for ~ t > 0 
\end{equation}

\begin{figure}
 \includegraphics[width=20pc]{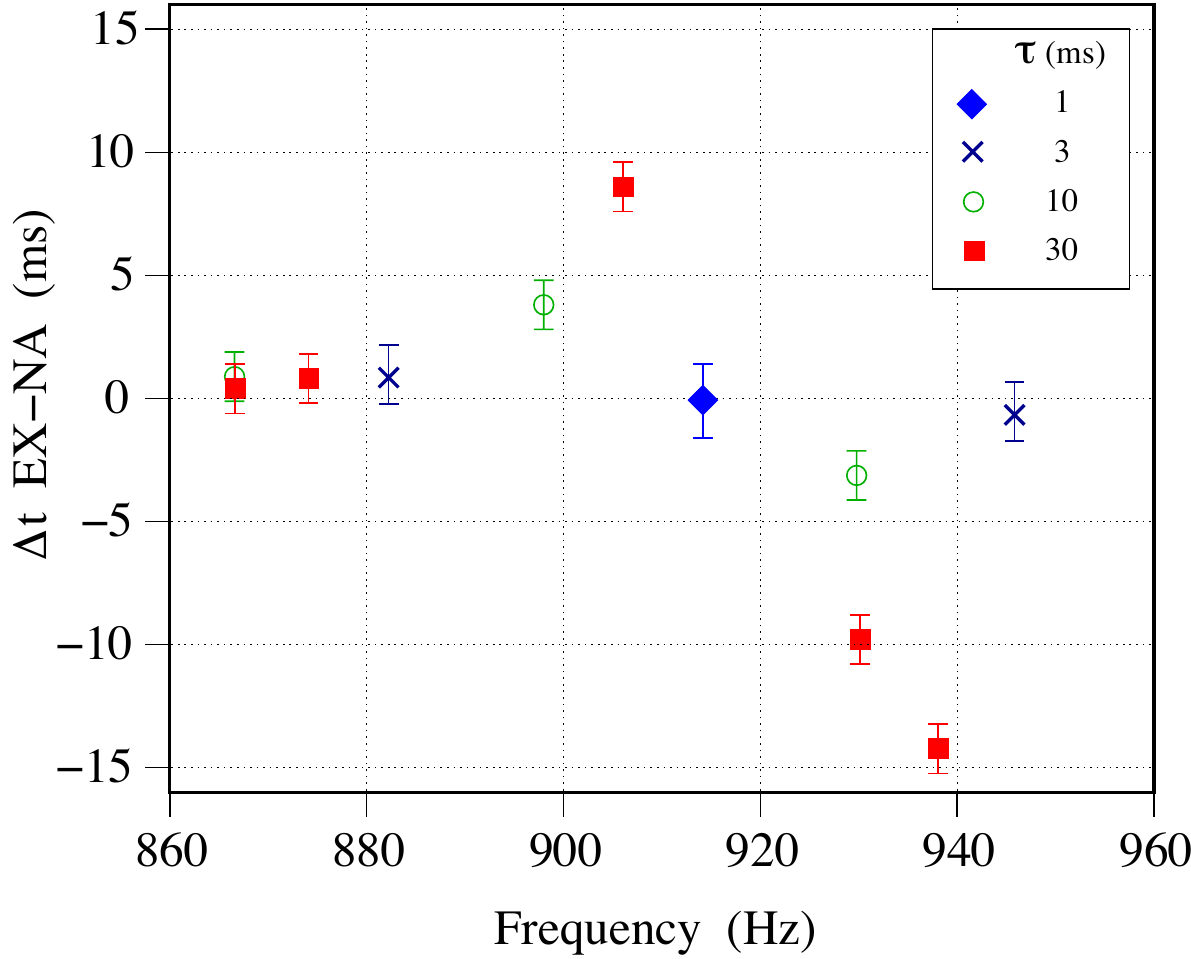}
\caption{ Offsets in the event time differences between Explorer and Nautilus for damped sinusoids with $\tau =$ 1,3,10 and 30 ms.  This systematic error varies between -15 and +10 ms.}
\label{fig:off_dten}
\end{figure}

 We summarize here the main results of this analysis, relevant for this IGEC2 search:
\begin{itemize}
\item  The shortest bursts ($\tau =1  ms$) excite the detectors in a fashion very close to that expected for delta pulses, confirming the predictions of analytical models, and providing further details and confidence.
\item   The detection efficiency of the antennas  remains good also for longer burst, provided that they carry a sufficient Fourier component in the detector bandwidth. If we compare pulses with the same $h_{rss}$,  waves with  $\tau >$ 10  ms often produce larger responses than the shorter ones, despite the use of the delta-matched filter.
\item  A large fraction of the timing uncertainty is usually due to a systematic offset that has a different dependence on signal frequency and $\tau$ in each antenna (see fig.\ref{fig:off_dten})
\item The width of the timing uncertainty (around the offset value) depends on the SNR of the signal and only slightly, if at all, on its shape.
\end{itemize}

\begin{figure}
 \includegraphics[width=20pc]{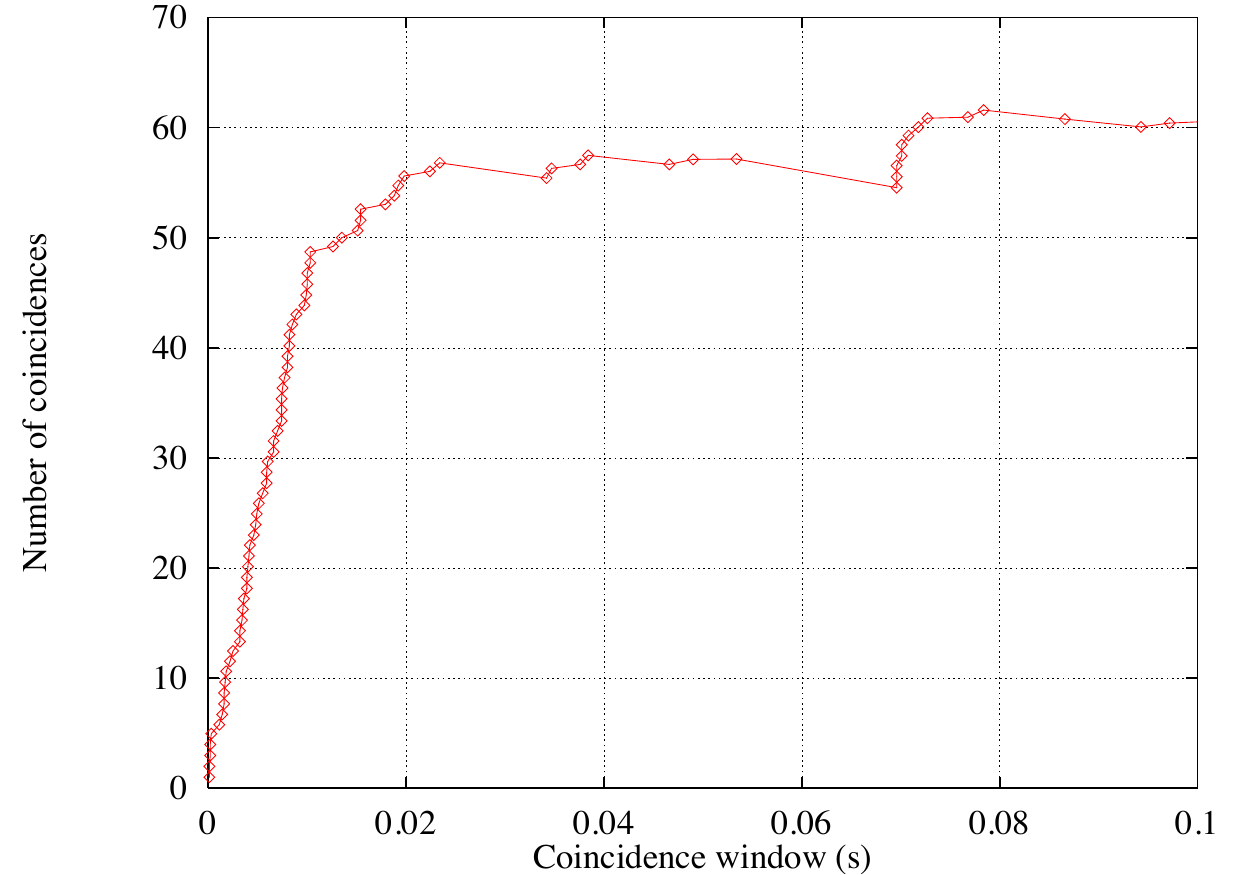}
\caption{Cumulative number of coincidence found in EXPLORER-NAUTILUS vs the absolute value of the event time difference, obtained by subtracting the theoretical cumulative of accidentals from the cumulative of events found in proximity of the injection times. The common excitations are software injections of damped sine waves with  $f=914 Hz$, $\tau = 1ms$.  About 95\% of the events  with $4.2 \le SNR \le 4.6$ are retrieved with a coincidence window of $\pm 25 ms$.}
\label{cumul}
\end{figure}

The goal of this study is to set a coincidence window as narrow as possible to minimize the background, while accommodating all these offsets and fluctuations in order to keep the detection efficiency high also for non delta-like excitations. To this aim, we found it useful to investigate the effect of these uncertainties on a pair of detectors, rather than on a single one.
As an example, we report studies carried out for the EXPLORER-NAUTILUS pair. Regarding the systematic time offset, fig.\ref{fig:off_dten} shows that, as the center frequency or the duration of the injected signal is changed, the offset difference can vary between -10 and +15 ms.
 As far as statistical fluctuations are concerned, fig.\ref{cumul}  shows the cumulative distribution of the detection time difference in the case of short signals. We can see that the first part of the curve can be well fitted by a gaussian with $\sigma_t$ = 9 ms, while for larger time differences there is a longer tail, containing about 5 -10\% of the events. We can conservatively state that 95\% of the coincidences is retrieved with a window of 25 ms.

As a consequence of these considerations, we have set the coincidence time window to 40 ms,
thus including the possible offsets deriving from non-$\delta$ waveforms, the statistical
fluctuations and the time of flight (up to $\sim$ 2 ms for the European antennas).
In the coincidences involving ALLEGRO data, the window is expanded to 60 ms, in order to take into account the larger maximum ($\sim$ 20 ms) time of flight of a signal between ALLEGRO and the other detectors.

\subsection{The choice of thresholds}

In a previous search \cite{IGEC2} that focused on triple coincidences,  we explored three different strategies for the choice of the thresholds: A) constant, equal SNR in all antennas, B) constant SNR in each antenna and higher threshold on the more sensitive detector C) constant, equal absolute amplitude ($H$) reconstructed by each antenna.

In the present analysis, 
thanks to the considerations detailed below, we were able to focus the analysis  on one criterion,  choosing, since the beginning,
 strategy B: higher SNR threshold for AURIGA, same SNR thresholds for the other three detectors. This allows,
on average, a better efficiency to different classes of signal waveforms.

\begin{figure}
\includegraphics[width=20pc]{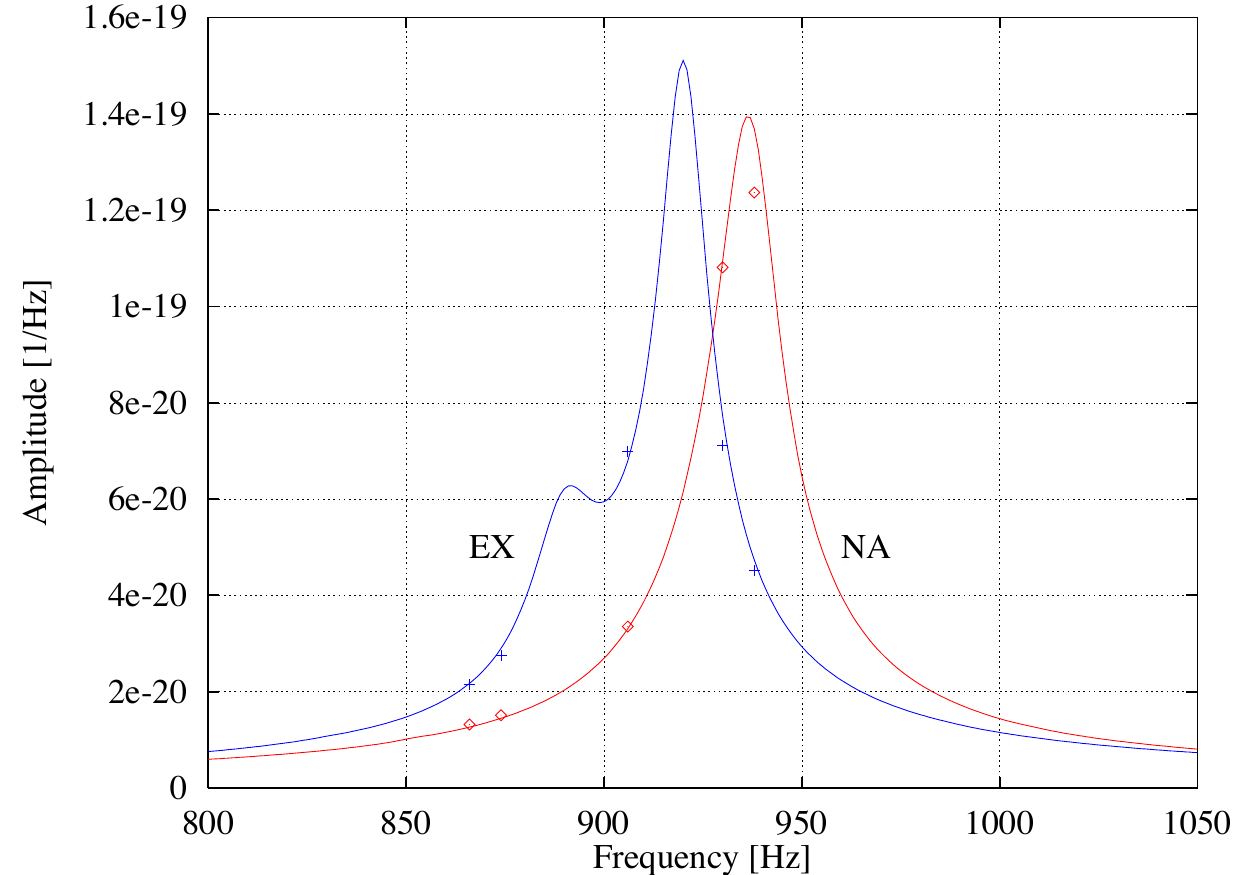}
\caption{ Responses of detectors to a damped sinusoid ($h_{rss} = 1 \cdot 10^{-18} Hz^{-1/2}$, $\tau= 30$ms).  The plot shows the  amplitude response of Explorer and of Nautilus vs the wave center frequency. Markers indicate values measured via software injections, while the lines show the behavior predicted by an analytical model.}
\label{fig:dampusoid}
\end{figure}

In fact, in the methodological study cited above \cite{virgobarre}, we also investigated how different detectors respond to the same, non-delta excitation. As an example we show in fig.\ref{fig:dampusoid} the responses of EXPLORER and NAUTILUS to damped sinusoids, linearly polarized and optimal oriented, with amplitude $h_{rss} = 1 \cdot 10^{-18} Hz^{-1/2}$ and decay time $\tau= 30$ ms, as a function of their central frequency $f_0$. 

\begin{figure}
 \includegraphics[width=20pc]{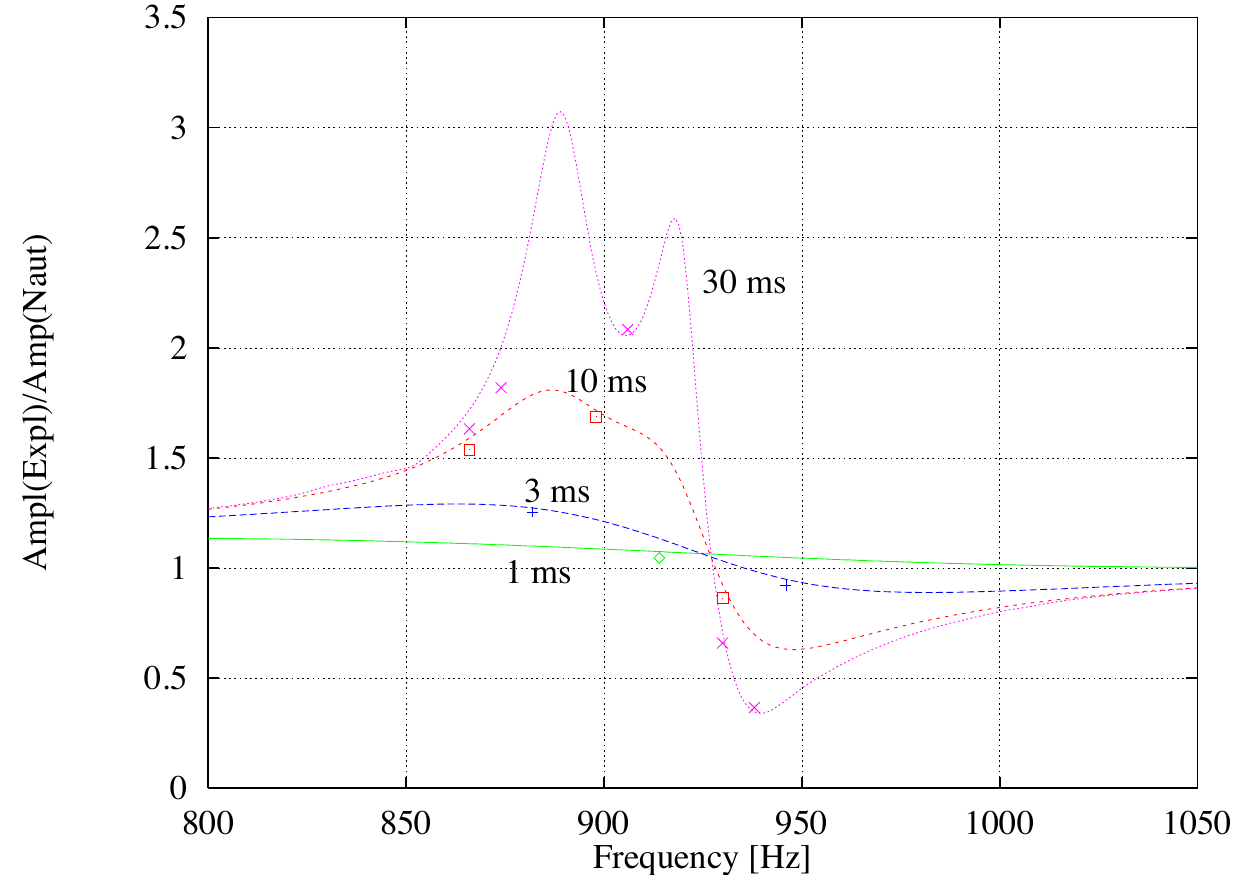}
\caption{Responses of detectors to damped sinusoids  with various duration ($\tau= $1,3,10,30 ms). The plot shows the ratio of the responses (Amplitude Explorer/ Amplitude Nautilus) vs the wave center frequency. 
Markers and lines as in fig.\ref{fig:dampusoid}.}
\label{fig:variatau}
\end{figure}

Analogously, fig.\ref{fig:variatau} shows the ratio of the responses (EXPLORER/NAUTILUS) for 4 different values of $\tau$:  t is evident that, while this ratio is close to unity for short bursts, approaching delta pulses, it can vary up to a factor 3 for longer signals,
since these two detectors, so similar under many aspects, do have narrow, not fully overlapping bandwidths.

These considerations clearly show that the same excitation can produce different responses in the detectors. Since we are searching small, near threshold signals, we chose not to implement criteria that select events by requiring equal or comparable amplitude response, because this would only select short bursts, \textit{a priori} rejecting longer signals for which the detectors are just as sensitive.
Therefore, it does not appear sensible to apply a threshold based on the absolute $H$ value of the candidate events. This leads us to discard criterion C, that anyhow was adopted for calculating upper limits, not a goal for this paper.
Both A and B criteria improve the efficiency to longer bursts with colored spectrum. Criterion B, instead, shows a better efficiency than A for short bursts, as this would give a higher SNR in the detector with better spectral sensitivity. Therefore in this search we adopted only criterion B.

\subsection{Background estimation and fine tuning of the amplitude thresholds}

The FARs were estimated with the usual time-shifting technique, i.e. creating a large number of replicas of the same event lists with the event times of each list changed by offsets. In the past, this was done by choosing one shift step and producing replicas with offsets that are multiples of such step.

The expected statistical distribution of accidental background can deviate from the Poisson distribution if clustering or non-stationarity affect the event lists. These deviations would go undetected if an uneffective time shifts is step chosen: large time shift steps do not correctly display possible correlation effects in nearby events, producing a poissonian behavior even if the actual background of accidental coincidences shows a more complex structure at small time shifts. This effect practically mimics a {\it decorrelation} of the data. On the other hand, a small shift step does not allow to probe a sufficiently large time lag, producing replicas that are not completely independent.

In order to avoid this effect, we used, in each background evaluation (see table \ref{tab:fa_snr}) a
series of 13 separate sets of evenly spaced time offsets, differing by the values of shift steps $\{T_{offset}\}$ =
\{0.12,0.36, 0.60, 0.84, 1.08, 1.32, 1.56, 1.80, 2.04, 2.28,
2.52, 2.76, 3.00\} s. The corresponding maximum relative time shifts between any pair of detectors result in the range $\pm$240 to $\pm$6000 s.
As an additional check, we verified the results of this procedure with other two alternative methods.
The first one consists in assigning a random time offset, picked from a uniform distribution to the event list of each detector.
With this method, the distribution of time shift differences between any pair of detectors is not uniformly distributed but concentrated around zero, thus better probing the zone around the zero-lag point. The second method is the one described in the appendix of \cite{IGEC2}. In this case the accidentals are analytically estimated from the number of events in each single detector found in random subdivisions of the overlap intervals. The results of both of these alternative methods agree, within statistical uncertainties, with those of the first technique, i.e. of the multiple time steps.

\begin{figure}
 \includegraphics[width=20pc]{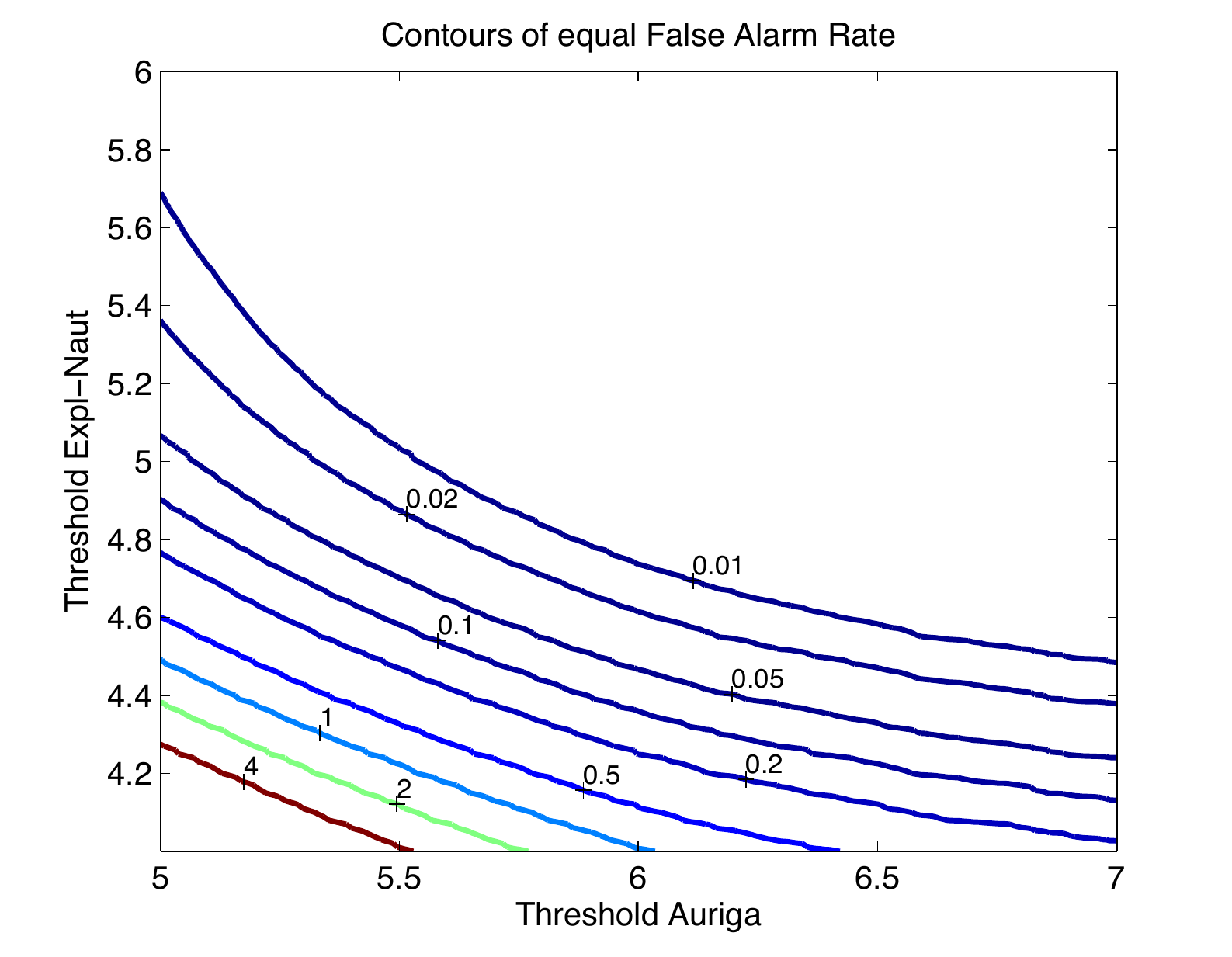}
\caption{Contour lines of the accidental rate (events/century) estimates in the AU-EX-NA (data stretch B)case.}
\label{figContour}
\end{figure}

\subsubsection{Threefold coincidences analysis} 

The target FAR  of 0.2 events/century for each of the four different
configurations of threefold coincidence was obtained by selecting 
the candidate events on the base of SNR threshold. 

The thresholds were tuned requiring the same SNR value for ALLEGRO, EXPLORER and NAUTILUS, and an higher SNR value for AURIGA, which has a noise level about 2 times smaller than the other detectors, thus
balancing the average sensitivities of the different detectors.

\begin{table}[h]
	\centering
			\begin{tabular}{|c|c|c|c|c|c|c|c|c|}\hline
			&\textbf{Overlap }& &\multicolumn{6}{c|}{\textbf{SNR threshold}} \\ \cline{4-9}

			\textbf{Configuration}& \textbf{time}  &&\multicolumn{3}{c|}{\textbf{Stretch A}}& \multicolumn{3}{c|}{\textbf{Stretch B}} \\  
			 &(days)  && \multicolumn{3}{c|}{\textbf{2005-2006}}& \multicolumn{3}{c|}{\textbf{2007}}\\ \hline 
	
			AL AU EX  & 361.8  && 4.3    &  6.88 &  4.3   &4.3    &  6.24 &  4.3   \\   \hline  
			AL AU NA & 390.6  && 4.3    & 7.1   &  4.3   &4.3    & 6.34   &  4.3  \\ \hline  
		  	AL EX NA  & 308.7  && 4.94  & 4.94  & 4.94  & 4.84 & 4.84   & 4.84 \\ \hline  
			AU EX NA & 301.9  && 6.88   & 4.2   &  4.2  &6.14   & 4.2     &  4.2 \\ \hline  		 	
		\end{tabular}
	\caption{ Time of simultaneous operation of the four threefold configurations and SNR threshold for the various  detectors . The different sets of values  in Stretch A and Stretch B were chosen in order to yield the same FAR  of 0.2/century in each search. }
	\label{tab:snr_cut}
\end{table}

The selection was done starting from the lowest possible SNR threshold, i.e. using all the exchanged data, and rising the threshold values till the target FAR was reached. Fig.\ref{figContour} shows as an example how the FAR varies with the threshold values in AURIGA and in the pair EXPLORER-NAUTILUS for which a common value was adopted.
This iterative process brought us to implement the SNR thresholds reported, for each configuration, in table \ref{tab:snr_cut}. 

We performed 12,006,000 time shifts per each of the 13 steps cited above.  

All 13 different statistical distributions of the four configurations of three detectors are well fitted by a poissonian.
The 52 mean values of the poissonian (i.e. the false alarms) are shown in table \ref{tab:fa_snr}; in the last row we report the mean and standard deviation per each detectors configuration.
The experimentally determined ratio between standard deviation and mean is about $2.5\cdot10^{-2}$, larger than expected from a Poisson counting statistics, which would give $6-7\cdot10^{-3}$.

For each threefold configuration, we built the overall distribution of the 13 ones, relative to different shift steps.
The predicted probabilities of 0,1,2 and 3 accidental counts per each detectors configuration are reported in table \ref{tab:tab_res_trip}.

The values in the first rows of each configuration are relative to the experimental background and were obtained calculating mean and standard deviation of the correspondent bins of 13 distributions each representing 12,006,000 trials. The second rows of each configuration show the probability values in the hypothesis of poissonian distributions having means as in the last row of table \ref{tab:fa_snr}.
The experimental and poissonian values are compatible within the uncertainties.

We notice that to calculate the experimental values we did not consider that the same time shift can be obtained using different steps, but we verified that the effect of this overlap among the 13 used distributions is negligible.

\subsubsection{Fourfold coincidences analysis} 
For the fourfold configuration, all the exchanged events of ALLEGRO, EXPLORER and NAUTILUS were used
(thresholds as in table \ref{tab:thr}), while AURIGA data were selected so as to reach the desired 0.2 false alarms per century. AURIGA threshold resulted  SNR = 4.96 for 
stretch A and   SNR = 4.57 for stretch B (2007 data).
As for the threefold case, also the analysis of the fourfold coincidence background was performed using the 13 time shift steps $\{T_{offset}\}$.
The total number of independent trials for each shift  was $11,094,161$. The last column of table \ref{tab:fa_snr} reports the measured false alarms for each $\{T_{offset}\}$ step. 

The last row of table \ref{tab:tab_res_trip} contains the experimental probabilities of
obtaining 0,1,2,3 coincidences by chance. These values can be compared with those in the second line in the row, relative to a Poisson distribution 
of mean $0.199\pm0.003$.

Similarly to the threefold coincidence analysis, the experimental probability is fully compatible with a poissonian distribution within the calculated errors.

\section{\label{sec:concl} Results of the Search and Conclusions}

When the confidential time offsets were finally exchanged, the true time analysis was performed
and no {\it real} (i.e. true time) coincidence was found in any of the five parallel searches that, based on our background estimate, yielded a total FAR of 1 event per century. 
We also looked for coincidences on all exchanged data, i.e. with the amplitude thresholds of the data exchange shown in table \ref{tab:snr_cut}.
We found no quadruple coincidence and 20 triples, well within the expected occurrence of accidentals ($\simeq$ 22, as reported in table \ref{tab:FA_all}). 
Since in this case the FAR is $\sim$ 14/year, this analysis cannot identify single GW candidates with a reasonable significance. In fact, we \textit{a priori} agreed that the only valid IGEC2 search for GWs was the former at FAR=1/century. 
Moreover, the total number of coincidences found is well within the expectations. Nonetheless, as we had \textit{a priori} decided, we make public \cite{web} these coincidences to allow possible further analysis with the use of fresh data from other detectors or astrophysical observations.
To allow a blind analysis of these data the published times of these coincidences have been shifted within $\pm$ 10 s, as performed internally by IGEC2.
The signals with lowest amplitudes that IGEC2 can detect have an $h_{rss}$ of the order of a few $10^{-20} Hz^{-1/2}$, which is larger by more than one order of magnitude with respect to the current LIGO-Virgo network sensitivity for signals in the same frequency band \cite{bursts5}.
Indeed, a simple calculation shows that, due to the long observation time, IGEC2 can set an upper limit on incoming g.w. flux that is twice better than that set by LIGO S5 run \cite{bursts5,LIGOS5}, but only for amplitudes $ h_{rss}> 10^{-19} Hz^{-1/2}$. 

In this paper we have reported on a search for coincident events on four resonant gravitational wave detectors, covering a data taking period of 17 month
that produced no candidate event within our choice of confidence level.
  Since ALLEGRO ceased operating in 2007,
 this is a final report for the four antennas array, but  the survey for a rare, highly energetic GW event does continue with  the three detectors  still in operation. Next searches will most likely be based on triggers from other observatories (neutrinos, X or gamma rays, etc.) due to the lower statistical robustness of a network of three antennas, all located in the same geographic area.

We believe that  the IGEC2 survey can complement the observation time of the LIGO-Virgo network in the next years, until the long baseline interferometers resume
their long term observations after the planned upgrades.

 \bibliographystyle{amsplain}

\begin{table*} [h]
	\centering
			\begin{tabular}{|c|c|c|c|c|c|c|}\hline
			& \multicolumn{6}{c|}{ {False Alarm Rates}} \\
		\textbf{Shift}	& \multicolumn{6}{c|}{\textbf{(events / century) }} \\ \cline{2-7}
		\textbf{(s)}  &{AL AU EX}& {{AL AU NA}}& {{AL EX NA}}& {{AU EX NA}}&& {{AU EX NA AL}}\\  \hline

0.120 & 0.204 & 0.234 & 0.215 & 0.203 && 0.200\\ \hline  
0.360 & 0.200 & 0.196 & 0.205 & 0.195 && 0.200\\ \hline  
0.600 & 0.195 & 0.199 & 0.206 & 0.196 && 0.194\\ \hline  
0.840 & 0.192 & 0.199 & 0.211 & 0.205 && 0.195 \\ \hline  
1.080 & 0.200 & 0.200 & 0.202 & 0.202 && 0.198\\ \hline  
1.320 & 0.204 & 0.199 & 0.200 & 0.207 && 0.205\\ \hline  
1.560 & 0.199 & 0.198 & 0.197 & 0.203 && 0.199\\ \hline  
1.800 & 0.191 & 0.198 & 0.196 & 0.200 && 0.199\\ \hline  
2.040 & 0.199 & 0.197 & 0.206 & 0.206 && 0.196\\ \hline  
2.280 & 0.192 & 0.197 & 0.198 & 0.201 && 0.199\\ \hline  
2.520 & 0.189 & 0.206 & 0.194 & 0.203 && 0.196\\ \hline  
2.760 & 0.197 & 0.203 & 0.194 & 0.210 && 0.201\\ \hline  
3.000 & 0.200 & 0.201 & 0.191 & 0.204 && 0.200 \\ \hline 
 & & & & & & \\[-3mm]
 \hline
mean, st.dev. & 0.197, 0.005
&0.202, 0.010& 0.201, 0.007 &0.203, 0.004 &&0.199, 0.003 \\ \hline
 \end{tabular}
	\caption{False Alarm Rates (FAR)  of the four coincidences analyses (three threefold and one fourfold), evaluated with about 12 millions trials for each of the 13 time shifts steps of the set  $\{T_{offset}\}$:  
In each search the thresholds were chosen according to the values of table \ref{tab:snr_cut}}
\label{tab:fa_snr}
\end{table*}

\begin{table*}
	\centering
			\begin{tabular}{|c|c|c|c|c|}\hline		
 & & & & \\[-3mm]
					 			\textbf{Configuration} &{\textbf{P(N=0)}}& {\textbf{P(N=1)}} & {\textbf{P(N=2)}}& {\textbf{P(N=3)}}\\ 

 & & & & \\[-3mm]
			\hline
AL AU EX & $0.998049 \pm 4.6\E{-5}$ & $(1.949 \pm 0.046)\E{-3}$ & $(2.02 \pm 0.49)\E{-6}$ & $<8\cdot 10^{-8}$ \\
 & $0.998049 \pm 4.7\E{-5}$ & $(1.949 \pm 0.047)\E{-3}$ & $(1.904 \pm 0.091)\E{-6}$ & $(1.241 \pm 0.088)\E{-9}$ \\
\hline
AL AU NA & $0.99784 \pm 1.0\E{-4}$ & $(2.15 \pm 0.10)\E{-3}$ & $(2.23 \pm 0.35)\E{-6}$ & $<8\E{-8}$ \\
 & $0.99784 \pm 1.0\E{-4}$ & $(2.19 \pm 0.10)\E{-3}$ & $(2.34 \pm 0.22)\E{-6}$ & $(1.69 \pm 0.24)\E{-9}$ \\
\hline
AL EX NA & $0.998299 \pm 5.7\E{-5}$ & $(1.700 \pm 0.057)\E{-3}$ & $(1.49 \pm 0.29)\E{-6}$ & $<8\E{-8}$ \\
 & $0.998299 \pm 5.7\E{-5}$ & $(1.700 \pm 0.057)\E{-3}$ & $(1.448 \pm 0.097)\E{-6}$ & $(8.24 \pm 0.83)\E{-10}$ \\
\hline
AU EX NA & $0.998325 \pm 3.4\E{-5}$ & $(1.674 \pm 0.034)\E{-3}$ & $(1.51 \pm 0.26)\E{-6}$ & $<8\E{-8}$ \\
 & $0.998325 \pm 3.4\E{-5}$ & $(1.674 \pm 0.034)\E{-3}$ & $(1.40 \pm 0.057)\E{-6}$ & $(7.85 \pm 0.48)\E{-10}$ \\
\hline
 & & & & \\[-3mm]
\hline
AL AU EX NA & $0.998402 \pm 2.4\E{-5}$ & $(1.598 \pm 0.024)\E{-3}$ & $(2.1 \pm 3.9)\E{-7}$ & $<9\E{-8}$ \\
 & $0.998403 \pm 2.4\E{-5}$ & $(1.595 \pm 0.023)\E{-3}$ & $(1.275 \pm 0.038)\E{-6}$ & $(6.79 \pm 0.30)\E{-10}$ \\
\hline
\end{tabular}

\caption{The first line of each configuration contains the experimental occurrence probabilities of 0,1,2 or
3 accidental coincidences calculated averaging the results of about 12 milions independent
trials for each of the 13 shifts steps of the set  $\{T_{offset}\}$.
The second rows of each configuration show the poissonian probability values in the hypothesis of  means as in the last row of table \ref{tab:fa_snr}. }
	\label{tab:tab_res_trip}
\end{table*}

\begin{table}[h]
	\centering
			\begin{tabular}{|c|c|c|}\hline
			& \textbf{Expected}& \textbf{Coincidences}\\ 
			\textbf{Configuration}& \textbf{accidentals}& \textbf{found}\\ 
			 &\textbf{(events)}&\textbf{(events)}\\ \hline
			AL AU EX & $4.29 \pm 0.01$ & 3\\ \hline
			AL AU NA & $5.15 \pm 0.01$ & 5\\ \hline
			AL EX NA & $10.23 \pm 0.01$ & 8\\ \hline
			AU EX NA & $2.34 \pm 0.01$ & 4\\ \hline  
& & \\[-3mm]
\hline
			AL AU EX NA & $(7.66\pm0.01)\E{-3}$ & 0\\ \hline 
			\end{tabular}
	\caption{Number of expected false alarms and of coincidences found in the search with minimum thresholds.}
	\label{tab:FA_all}
\end{table}

\end{document}